\newcommand{\bef}{\begin{figure}}
\newcommand{\eef}{\end{figure}}

\newcommand{\be}{\begin{equation}}
\newcommand{\ee}{\end{equation}}
\newcommand{\bea}{\begin{eqnarray}}
\newcommand{\eea}{\end{eqnarray}}
\documentclass{PoS}
\title{Unfolding of event-by-event net-charge distributions in heavy-ion collisions}
\ShortTitle{Unfolding of event-by-event net-charge distributions in heavy-ion collisions}

\author{\speaker{P. Garg}\\
        Department of Physics, Banaras Hindu University, Varanasi-221005, India	\\
        E-mail: \email{prakhar@rcf.rhic.bnl.gov}}
\author{D. K. Mishra, P. K. Netrakanti, A. K. Mohanty\\
       Nuclear Physics Division, Bhabha Atomic Research Center, Mumbai 400094, India\\}
\author{B. Mohanty\\
        School of Physical Sciences, National Institute of Science Education and Research, Bhubaneswar 751005, India\\
        E-mail: \email{bedanga@niser.ac.in}}
\abstract{
An unfolding method, based on Bayes theorem is presented to obtain true event-by-event net-charge multiplicity distribution from a corresponding measured distribution, which is subjected to detector artifacts. The unfolding is demonstrated to work for widely varying particle production mechanism, beam energy and collision centrality. Further  the necessity of taking into account the detector effects is emphasized  before comparing the experimental measurements to the theoretical calculations, particularly in case of higher moments. The advantage of this approach being that one need not construct new observable to cancel out detector effects which loose their ability to be connected to physical quantities calculable in standard theories.}

\FullConference{$8^{th}$ International Workshop on Critical Point and Onset of Deconfinement\\
		March 11 - 15, 2013\\
		Napa, California, USA}

\begin{document}

\section{Introduction}

It has been suggested that the higher moments of fluctuations are very sensitive to the proximity of critical point as they have strong dependence on correlation length~\cite{Stephanov:2008qz} and they are related to the susceptibility of the system~\cite{Bazavov:2012jq,Cheng:2008zh}. The higher moments of event-by-event distribution of conserved quantities like net-charge, net-baryon and net-strangeness have been important observables to characterize the system formed in heavy ion collision experiments~\cite{Aggarwal:2010wy}. Further suggestions are made to explore the QCD phase transition and freeze out  conditions in heavy ion collisions, using the higher moments~\cite{Karsch:2010ck, Friman:2011pf}.    

Most of the experimental measurements depend on the detector acceptance, particle counting efficiency and other background effects~\cite{Abelev:2008ab}. Some of them are very difficult to exclude for an observable which is based on an event-by-event analysis. Therefore the experimentally measured event-by-event distributions are shown without taking care of these corrections~\cite{Stephanov:2008qz,Aggarwal:2001aa,Adcox:2002mm,BraunMunzinger:2011dn}. These corrections are applied on an average basis to correct the particle yields in heavy ion collision experiments~\cite{Abelev:2008ab}, but to apply these corrections on an event-by-event observable is not trivial. Hence, comparing uncorrected event-by-event observables with the theoretical observables should be done carefully as it may lead to different physics conclusions.     

Although, some observables have been constructed in order to cancel out the detector effects to first order~\cite{Mrowczynski:1999un,Voloshin:1999yf,Bialas:2007ed,Pruneau:2002yf}, but while making these constructs one may loose the ability to compare them to the theoretically calculated quantities. Therefore, to compare higher moments of multiplicity distributions with the theoretical results, one should consider the experimental artifacts.

In the present work, an approach based on bayesian theorem of probability is demonstrated to work successfully to remove the detector artifacts on an event-by-event basis. Such a method has some constraints in terms of proper detector modeling and a large event-by-event multiplicities.

\section{Event generators}

Two different event generators are used to explore our proposal. For the present study, HIJING ~\cite{Gyulassy:1994ew} (version 1.37) and THERMINATOR~\cite{Kisiel:2005hn} (version 2.0) event generators provide the facility to incorporate  different particle production mechanism. Using these event generators, net-charge distributions are obtained within the pseudo-rapidity  of -0.5 to 0.5 and transverse momentum range of $0.2<p_{T}<2.0 GeV/c$ with in full azimuthal coverage. The average charge particle counting efficiency is taken to be 65\%, derived from the charged pion efficiency, as is given in Ref.~\cite{Abelev:2008ab}. Further to demonstrate our proposal at different energies, HIJING is also used at $\sqrt{s_{\mathrm {NN}}}$ = 27, 39, 62.4, 130 and 200 GeV for most central Au+Au events.

\section{Bayes method for the unfolding of distributions}
The RooUnfoldBayes~\cite{Agostini:1995yr} algorithm of RooUnfold package~\cite{Adye:2011} is used to demonstrate the present proposal. The algorithm based on Bayes theorem of probability is used to show that the true distributions can be reconstructed from the distributions which are affected by systematic biases and detection efficiency.  

To demonstrate it, 5M Au+Au collision events are generated for each centrality bin using HIJING for 19.6 GeV and THERMINATOR for 200 GeV. For each event positive $(N^{+})$ and negative $(N^{-})$ charge particles are selected and a net-charge distribution $(\Delta N=N^{+}-N^{-})$ is constructed on an event-by-event basis.

Now, to mimic the experimental situation, individual $N^{+}$ and $N^{-}$ are smeared with a Gaussian function of width 10\%, and the mean value corresponding to the average efficiency of 65\% as obtained from the parametrization of the $p_{T}$ dependent efficiency for charged pions from STAR experiment~\cite{Abelev:2008ab}. Afterwards, these smeared distributions are used to construct the net-charge distribution, we will call it the \textit{measured distribution}.

\bef[h]
\begin{center}
\includegraphics[scale=0.6]{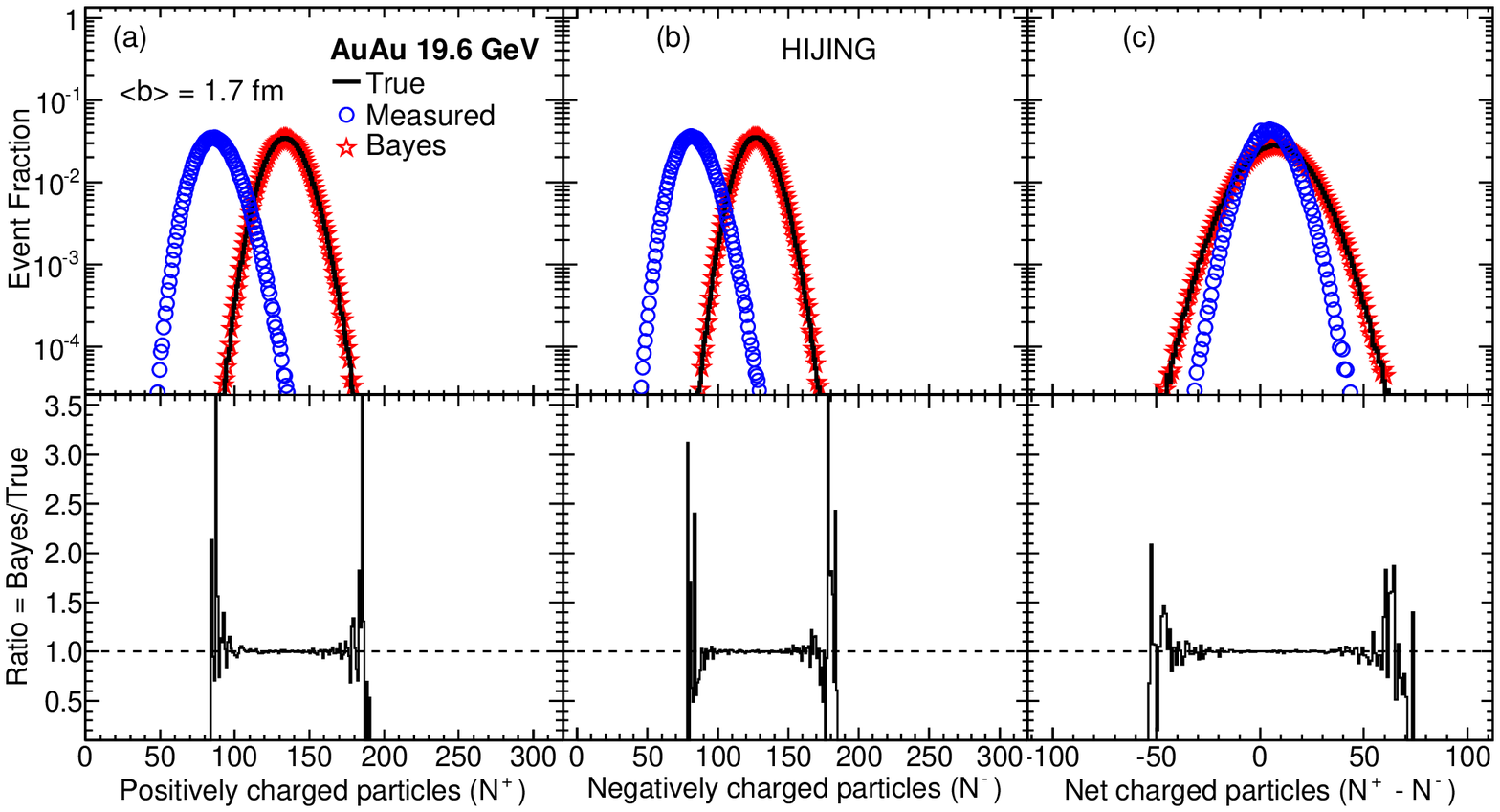}
\caption{(Color online) Top panel: Event-by-event distribution of positive, negative and
net-charge (denoted as ``True'', solid line) in Au+Au collisions for impact parameter $b$
= 1.7 fm at $\sqrt{s_{\mathrm {NN}}}$ = 19.6 GeV from HIJING event generator. Also
shown are the corresponding distributions after applying acceptance and efficiency effects
as discussed in the text (denoted as ``Measured'', open circles). The unfolded
distributions are shown as red stars and denoted as ``Bayes''. Bottom panel: Shows the
ratio of the unfolded to the True distributions.}
\label{fig1}
\end{center}
\eef

The measured distribution of net-charge is unfolded with response matrix obtained from the training procedure using iterative Bayes theorem.  The present study uses the optimal value of $4$ for the regularization. True, measured and unfolding are performed in a way to eliminate the finite centrality bin-width effect. The moments of net-charge distributions are derived using cumulant method \cite{luo}.

\section{Results and Discussions}

The true, measured and unfolded distributions for positive charge (panel a), negative charge (panel b) and net charge (panel c) are shown in Figure~\ref{fig1}.
These distributions are for most central events corresponding to an average impact parameter of 1.7 fm of Au+Au collisions from HIJING at 19.6 GeV on an event by event basis.
Solid lines, open circles and red stars represent the true distributions, measured distributions and the unfolded distributions respectively, for all the cases. It is evident that for all the cases, the true distributions are reproduced from the measured distributions, using the unfolding technique. Also the ratios, presented in the bottom panel suggest that the unfolding procedure is able to get back the true distribution. 

We have seen the variation of mean, sigma, skewness and kurtosis as a function of centrality ($N_{part}$), obtained from the net-charge distributions in Au+Au collisions at $\sqrt{s_{\mathrm {NN}}}$ = 19.6 GeV for true, measured and unfolded distributions~\cite{Garg:2012nf}. The moments computed from unfolded distributions and true distributions were found in good agreement. Further, the ratios of unfolded to true moments were close to unity. It suggests that the unfolding method reproduced the results of true distribution from the measured distributions. Same conclusions can be drawn from Fig.~\ref{fig4} where $\sigma^2/M$, $S\sigma$ and $\kappa\sigma^2$ are drawn as a function of $N_{part}$.

\bef[h]
\begin{center}
\includegraphics[scale=0.7]{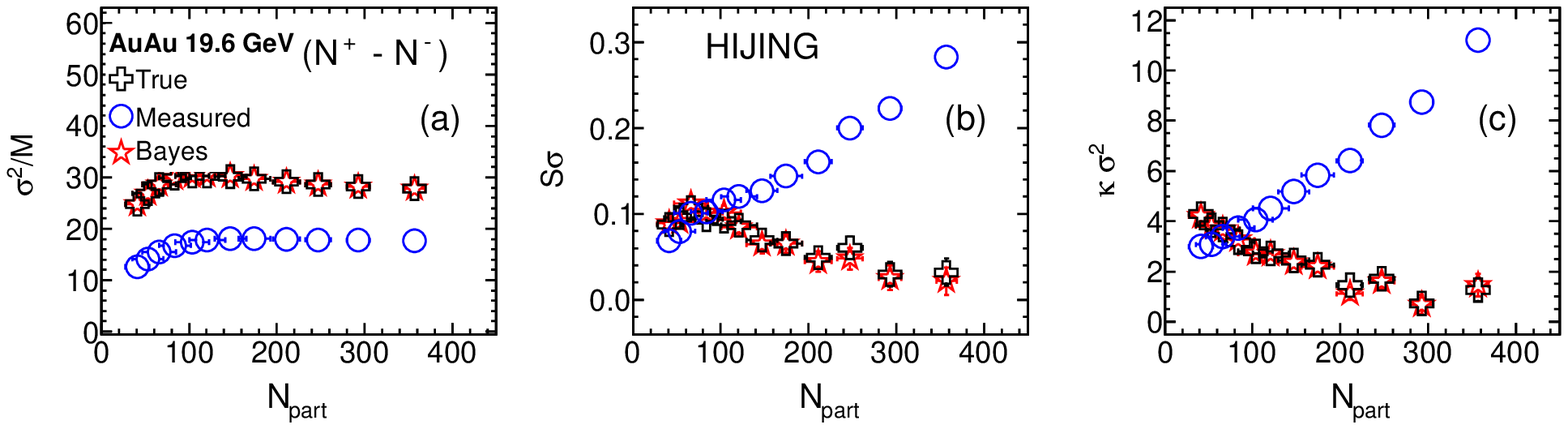}
\caption{(Color online) Ratio (panel a) and product of moments (panel (b) and (c)) of net-charge distributions in Au+Au collisions at $\sqrt{s_{\mathrm {NN}}}$ = 19.6 GeV from HIJING event generator. The
results are for the True, measured  and Bayes unfolded distributions as a function of $N_{\rm
{part}}$.}
\label{fig4}
\end{center}
\eef

It is observed (Fig.~\ref{fig6}) that the dependence  of $\kappa\sigma^2$ and $S\sigma$ is very different for true and smeared distributions, as a function of $N_{part}$. It implies that any physics conclusion associated with the variation of $S\sigma$ and $\kappa\sigma^{2}$ with $N_{part}$ for net-charge distributions could be highly misleading. However, the results for measured distributions can be unfolded nicely to get back the results of true distributions.

\bef[h]
\begin{center}
\includegraphics[scale=0.7]{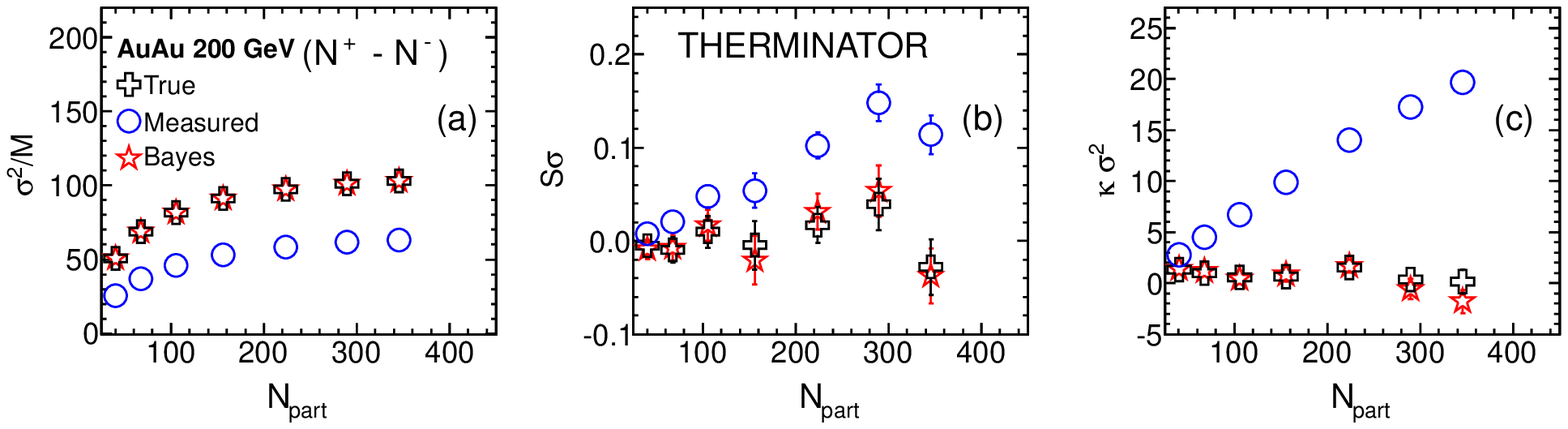}
\caption{(Color online) Product of moments of net-charge distributions in Au+Au collisions
at $\sqrt{s_{\mathrm {NN}}}$ = 200 GeV from THERMINATOR event generator. The results are
for the True, measured and Bayes unfolded distributions as a function of $N_{\rm
{part}}$.}
\label{fig6}
\end{center}
\eef

Another event generator, THERMINATOR is also used to check the validity of proposed bayesian approach.
Figure~\ref{fig6} shows the $\sigma^2/M$, $S\sigma$ and $\kappa\sigma^2$ of net-charge distributions from the true, measured and unfolded distributions at  $\sqrt{s_{\mathrm {NN}}}$ = 200 GeV, as a function of $N_{part}$. Here also the ratio and products of moments from unfolded distributions are reproduced as true distributions up to a good extent, suggesting that the method proposed in this paper works equally well for parent distributions produced from very different particle production mechanisms as well as over a wide range of beam energies. 
Besides 19.6 GeV, HIJING is used for $\sqrt{s_{\mathrm {NN}}}$ = 27, 39, 62.4, 130 and 200 GeV with the same procedure. In this energy dependence study only 0-5\% central Au+Au events are used.

In Figure~\ref{fig10}, $\sigma^2/M$, $S\sigma$ and $\kappa\sigma^2$ are drawn as a function of $\sqrt{s_{\mathrm {NN}}}$ for true, measured, and unfolded distributions. Here also a good agreement is found between True and Unfolded moments. This demonstrate that the proposed method works over a wide range of energies as well.

\bef[h]
\begin{center}
\includegraphics[scale=0.6]{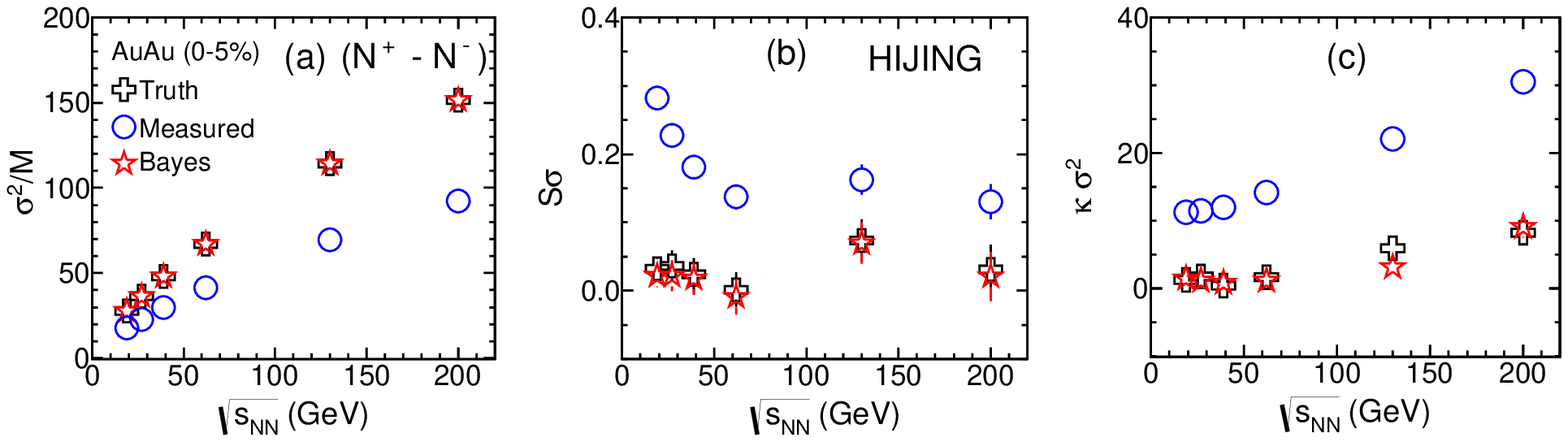}
\caption{(Color online) Product of moments of net-charge distribution in 0-5\% Au+Au collisions
as a function of $\sqrt{s_{\mathrm {NN}}}$ from HIJING event generator. The results are
for the True, measured and Bayes unfolded distributions.}
\label{fig10}
\end{center}
\eef



\section{Summary} 
The Bayesian unfolding method is successfully demonstrated to unfold back the measured distributions, which are subjected to detector effects like finite particle counting efficiencies. The centrality dependent study for moments and their product and ratios is carried with HIJING and THERMINATOR at 19.6 GeV and 200 GeV respectively. It is observed that the detector effects can modify the results significantly and these effects can be removed by bayesian unfolding method.
Also, for wide range of energies ($\sqrt{s_{\mathrm {NN}}}$=19.6, 27,39, 62.4, 130 and 200 GeV), a good agreement is found between true and unfolded moments, $\sigma^{2}/M$, $S\sigma$ and $\kappa\sigma^{2}$. However, there are limitations in terms of proper modeling of the detector response. Also it requires high multiplicity and large event statistics for building better response matrix. Although main advantage of Bayesian approach is that, one don't have to construct new observables to cancel out the detector effects. Further, more details of the present work can be found in Ref.~\cite{Garg:2012nf}.

\section{Acknowledgements}
Financial assistance from the Department of Atomic Energy, Government of India is gratefully acknowledged. BM is supported by the DST Swarna Jayanti project fellowship. PG acknowledges financial support from CSIR, New Delhi, India. This work is carried out using the NPD-BARC cluster facility at HBNI.

 \end{document}